\documentclass[notoc,hyper,letterpaper]{JHEP3}

\usepackage{epsfig}
\usepackage{amsbsy}
\usepackage{varioref}
\usepackage{pifont}

\title{Generalized Weinberg Sum Rules\\
 in Deconstructed QCD}

\author{R. Sekhar Chivukula\\
Department of Physics and Astronomy, Michigan State University\\
East Lansing, MI 48824, USA\\
	E-mail: \email{sekhar@msu.edu}}
	
\author{Masafumi Kurachi\\
Department of Physics, Nagoya University\\
Nagoya 464-8602, Japan\\
	E-mail:\email{kurachi@eken.phys.nagoya-u.ac.jp}}

\author{Masaharu Tanabashi\\
Department of Physics, Tohoku University\\
Sendai 980-8578, Japan\\
	E-mail:\email{tanabash@tuhep.phys.tohoku.ac.jp}}

\abstract{
Recently, Son and Stephanov have considered an ``open moose'' as
a possible dual model of a QCD-like theory of chiral symmetry breaking. In this note
we demonstrate that although the Weinberg sum rules are satisfied in any
such model, the relevant sums converge very slowly and in a manner unlike QCD.
Further, we show that such a model satisfies a set of generalized sum rules. These
sum rules can be understood by looking at the operator product expansion for the correlation
function of chiral currents, and correspond to the absence of low-dimension gauge-invariant
chiral symmetry breaking condensates. These results imply
that, regardless of the couplings and $F$-constants chosen, the open moose is not the dual
of any QCD-like theory of chiral symmetry breaking. We also show that the generalized sum rules
can be ``solved'',  leading to a compact expression for the difference of vector- and axial-current correlation functions. This
expression allows for a simple formula for  the $S$ parameter ($L_{10}$),  which
implies that $S$ is always {\it positive} and of order one in any (unitary) open linear moose model. 
Therefore the $S$ parameter is positive and order one  in any ``Higgsless model''  based on the continuum limit of a linear moose regardless of the warping or position-dependent gauge-coupling chosen.
}

\keywords{Dimensional Deconstruction, Chiral Symmetry Breaking, Sum Rules}
\preprint{
{MSUHEP-040308} \\
{DPNU-04-05} \\
{TU-712}
}

\begin{document}



Recently, motivated by models of hidden local symmetry \cite{Bando:1985ej,Bando:1985rf,Bando:1988ym,Bando:1988br,Harada:2003jx}, gauge/gravity duality \cite{Maldacena:1998re,Gubser:1998bc,Witten:1998qj,Aharony:1999ti}, and dimensional deconstruction  \cite{Arkani-Hamed:2001ca,Hill:2000mu}, Son and Stephanov \cite{Son:2003et} have considered an ``open moose" as a possible dual model of a QCD-like theory with chiral symmetry breaking. The model, shown diagrammatically (using ``moose notation'' \cite{Georgi:1986hf,Arkani-Hamed:2001ca}) in Fig. \ref{fig:Moose1}, incorporates $K+1$ nonlinear $(SU(2)\times SU(2))/SU(2)$ sigma models in which the global symmetry groups in adjacent non-linear sigma models are identified through $K$ $SU(2)$ gauge groups. The far left- and right-handed symmetry groups (carried by $\Sigma^1$ and $\Sigma^{K+1}$ respectively) remain ungauged, and are identified with the (approximate) chiral $SU(2)_{L,R}$ of QCD. Parity symmetry is naturally incorporated by identifying the gauge-couplings and $F$-constants under reflection, $g_i = g_{K+1-i}$ and $F_i=F_{K+2-i}$.

\EPSFIGURE[h]{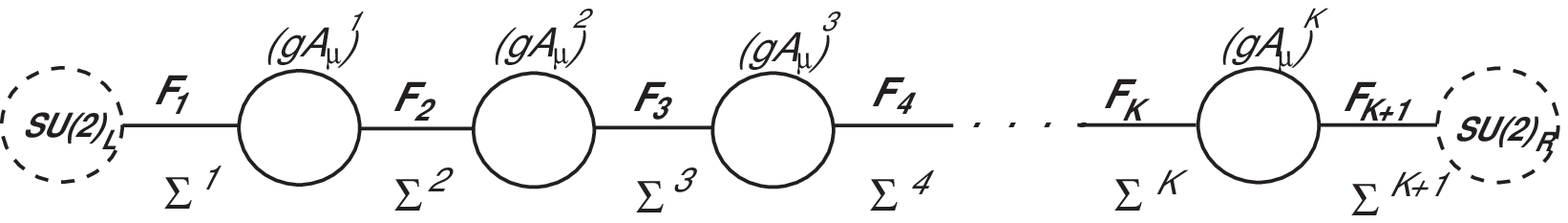,width=0.9\textwidth}
{Moose diagram for model.
\label{fig:Moose1}}

The identification of this model as a possible dual to a QCD-like theory rests principally on two features. First, the model has the right basic symmetry properties of QCD, including parity, the correct global symmetry breaking pattern ($SU(2)_L \times SU(2)_R \to SU_V(2)$), and a tower of vector and axial vector meson states. More remarkably, for an arbitrary set of parity-symmetric couplings $g_i$ and $F$-constants $F_i$, Son and Stephanov  \cite{Son:2003et} have shown that the model automatically respects the Weinberg sum rules \cite{Weinberg:1967kj} expected to be valid in QCD.

In this letter, we investigate the difference of vector-  and axial-current correlation functions in an arbitrary ``open moose'' model. By considering the asymptotic behavior of this difference of correlation function, we derive $K$ independent sum rules\footnote{After the completion of this work, we 
learned that eqn. (\protect\ref{result}), the generalized sum rules
(eqn. (\protect\ref{gsr})) which derive from it, and the interpretation of the sum rules in terms of the
absence of chiral-symmetry breaking condensates has been derived independently
by Hirn and Stern  \protect\cite{Hirn:2004ze}.}  which are automatically satisfied in any such model. The first two of these generalized sum rules are the Weinberg sum rules previously demonstrated by Son and Stephanov. We demonstrate that, unlike the behavior expected in QCD, the first and second sum rules are typically not well satisfied unless one includes all of the spin-1 resonances present in the model. That is, although the Weinberg sum rules are always formally satisfied in any ``open moose'' theory, no truncation of such a theory to the first few resonances can provide an approximation of QCD. 

Furthermore, relating our results to the operator product expansion \cite{Shifman:1979bx,Knecht:1998ts}, we show that the extra sum rules correspond to the absence of low-dimension gauge-invariant chiral symmetry breaking condensates. This absence is a direct result of the ``collective breaking'' of the chiral symmetry implicit in these models \cite{Arkani-Hamed:2001nc,Arkani-Hamed:2002pa}. These results imply that, regardless of the gauge couplings and $F$ constants chosen, the open moose is not the dual of any QCD-like theory. 

Finally, using these
results, we derive a simple expression for the difference between the vector and axial current correlation functions.  This expression allows us to evaluate the  $S$-parameter ($L_{10}$) and the analog of the $\pi^+$ - $\pi^0$ mass difference, and show that they are always positive and that the $S$ parameter
will be of order one in any unitary model.

The open moose is described by the following Lagrangian \cite{Son:2003et}
\begin{equation}
{\cal L} = {1\over 4} \sum^{K+1}_{k=1} F^2_k\,{\rm Tr}|D_\mu \Sigma^k|^2 
- {1\over 2}\sum_{k=1}^K \,{\rm Tr}(F^k_{\mu\nu})^2~.
\end{equation}
Here the covariant derivatives are defined as
\begin{equation}
D_\mu \Sigma^k = \partial_\mu \Sigma^k - i(gA_\mu)^{k-1} \Sigma^k
+ i\Sigma^k (g A_\mu)^{k}~,
\end{equation}
where we denote $g_k A^k_\mu \equiv (g A_\mu)^k$, the gauge field $A_\mu = A^a_\mu \tau^a/2$
(where the $\tau^a$ are the conventional Pauli matrices), and it is understood that 
the groups at the ends of the moose are not gauged: $g_0=g_{K+1} \equiv 0$ and 
$A^0_\mu = A^{K+1}_\mu \equiv 0$. The $K+1$ nonlinear sigma model fields $\Sigma^k(x)$
transform as
\begin{equation}
\Sigma^k \rightarrow U_{k-1}(x)\Sigma^k U^\dagger_k(x)~,
\end{equation}
where $U_{0}$ and $U_{K+1}$ are identified with the {\it global} transformations
$L,R \in SU(2)_{L,R}$. The $K$ gauge symmetries are spontaneously broken, yielding $K$
massive vector mesons which are to be identified with the vector mesons of QCD, and the
global symmetries $SU(2)_L \times SU(2)_R$ are broken to their diagonal subgroup, resulting
in a set of massless Goldstone bosons which are identified with the pions of QCD.

As usual, we may parameterize the fields $\Sigma^k$ by
\begin{equation}
\Sigma^k(x) = \exp\left({2i\pi^k(x)\over F_k}\right)~,
\end{equation}
of which $K$ linear combinations of the $\pi^k$ are ``eaten'' by
the gauge fields. The combination of sigma model fields
\begin{equation}
\Sigma = \Sigma^1 \Sigma^2 \ldots \Sigma^{K+1}~,
\end{equation}
transforms only under the the global symmetry groups
\begin{equation}
\Sigma \rightarrow L \Sigma R^\dagger~,
\end{equation}
and therefore the linear combinations
\begin{equation}
\pi = F\, \sum_{k=1}^{K+1} {\pi^k\over F_k};\ \ \ \ \ {1\over F^2} = \sum_{k=1}^{K+1} {1\over F^2_k}
\end{equation}
are the Goldstone bosons of chiral symmetry breaking, and $F$ is to be identified with $f_\pi$
in QCD. The vector-boson mass matrix may be computed directly, and is the tridiagonal matrix
\begin{equation}
{\tiny
M^2 = {1\over 4}
\left(
\begin{array}{c|c|c|c|c|c}
g^2_1(F^2_1+F^2_2)& -g_1 g_2 F^2_2 & & &  \\ \hline
-g_1 g_2 F^2_2  & g^2_2(F^2_2+F^2_3) & -g_2 g_3 F^2_3 & &   \\ \hline
 & -g_2 g_3 F_3^2 & g^2_3(F^2_3+F^2_4) & -g_3 g_4 F^2_4 &  \\ \hline
 & & \ddots & \ddots & \ddots &   \\ \hline
  & & & -g_{K-2} g_{K-1} F^2_{K-1} & g^2_{K-1}(F^2_{K-1}+F^2_{K}) & -g_{K-1} g_{K} F^2_K \\ \hline
 & & & & -g_{K-1} g_K F^2_K & g^2_K(F^2_K+F^2_{K+1})\\
\end{array}
\right)
}
\label{massmatrix}
\end{equation}
The eigenvectors $|\hat{l}\rangle$ of this matrix satisfy equation $M^2 |\hat{l}\rangle
= m^2_{\hat{l}} | \hat{l}\rangle$, with  $m^2_{\hat{l}}$ the masses of the vector mesons.
The mass eigenstate fields $A^{\hat{l}}_\mu$ ($\hat{l} =1,\ldots,K$) are then given by
\begin{equation}
A^{\hat{l}}_\mu = \sum_k  A^k_\mu \langle k | \hat{l} \rangle ~,
\end{equation}
where, for convenience, the expansion coefficients are written in Dirac
notation and satisfy the reality conditions $\langle k | \hat{l} \rangle^* = \langle \hat{l} | k \rangle
= \langle k | \hat{l} \rangle$ and the usual completeness conditions
\begin{equation}
\sum_{\hat{l}} \langle i | \hat{l} \rangle \langle \hat{l} | j \rangle = \delta_{ij}\ \ \&
\ \ 
\sum_k \langle \hat{l} | k \rangle \langle k | \hat{j} \rangle = \delta_{\hat{l}\hat{j}}~.
\end{equation}
Using the parity symmetry of the open moose we can classify the pions as pseudoscalars, and we find that the spin-1 spectrum alternates between vector and axial vector mesons.

Consistent with Son and Stephanov, we define the vector meson decay constants
\begin{equation}
\langle 0|J^a_{V\mu}(0) | A^{b\hat{l}}\rangle = g_{\hat{l} V} \delta^{ab} \varepsilon_\mu
\ \ \& \ \ 
\langle 0|J^a_{A\mu} (0)| A^{b\hat{l}}\rangle = g_{\hat{l} A} \delta^{ab} \varepsilon_\mu~,
\end{equation}
where $J^a_{{V,A}\,\mu}$ are the conventionally normalized vector and axial vector isospin currents.
Direct computation \cite{Son:2003et} then yields
\begin{equation}
g_{\hat{l}\,{V,A}}={F^2_1 g_1 \over 4} [\langle 1|\hat{l} \rangle \pm \langle K | \hat{l} \rangle]~,
\label{vectorcouplings}
\end{equation}
where we have used parity symmetry to set $g_1=g_K$ and $F_1 = F_{K+1}$. By parity symmetry, $g_{\hat{l}V}$ ($g_{\hat{l}A}$) vanishes for even (odd) $\hat{l}$.
Current conservation then implies that the correlation function of vector or axial currents may be
written in the form
\begin{equation}
\langle J^a_{{V,A}\, \mu}(q)\, J^b_{{V,A}\,\nu}(-q) \rangle = \delta^{ab}\,
(-q^2 \eta_{\mu\nu}+q_\mu q_\nu) \Pi_{V,A}(-q^2)~.
\end{equation}
In the open moose at tree level, one finds \cite{Son:2003et}
\begin{equation}
\Pi_V(Q^2) = \sum_{\hat{l}} {g^2_{\hat{l}V} \over m^2_{\hat{l}} (m^2_{\hat{l}} + Q^2)}~,
\end{equation}
and
\begin{equation}
\Pi_A(Q^2) = {F^2\over Q^2}+\sum_{\hat{l}} {g^2_{\hat{l}A} \over m^2_{\hat{l}} (m^2_{\hat{l}} + Q^2)}~.
\end{equation}

Of particular interest in the study of chiral symmetry breaking is the difference between the vector
and axial vector correlation functions
\begin{equation}
\Pi_{V-A}(Q^2)=\Pi_V(Q^2) - \Pi_A(Q^2) = -\,{F^2\over Q^2} + 
\sum_{\hat{l}} {{g^2_{\hat{l}V} -g^2_{\hat{l}A}}\over m^2_{\hat{l}} (m^2_{\hat{l}} + Q^2)}~.
\label{diffcorrelators}
\end{equation}
This correlation function can be written
in a surprisingly compact form. Using eqn. (\ref{vectorcouplings}), 
the numerator of the sum becomes
\begin{equation}
g^2_{\hat{l}V} - g^2_{\hat{l}A} = {F^4_1\, g^2_1\over 4}\, \langle 1 | \hat{l}\rangle \langle K | \hat{l} \rangle
= {F^4_1\, g^2_1\over 4}\, \langle 1 | \hat{l}\rangle \langle \hat{l} | K \rangle~.
\end{equation}
Then, using the completeness relations, we find that the difference of correlation functions
may be written entirely in terms of the vector meson mass matrix $M^2$
\begin{equation}
\Pi_{V-A}(Q^2) = -\,{F^2\over Q^2} + {F^4_1\, g^2_1\over 4}\, \left< 1 \left|
{1\over {M^2 ( Q^2\cdot {\cal I} +M^2)} }\right| K \right>~.
\label{result}
\end{equation}
This result
forms the basis of our discussion of the open moose model\footnote{In general, without
assuming parity symmetry, one obtains a result for the correlation function of left-
and right-handed currents; for $\Pi_{LR}$, there is an overall factor of $1/4$ and the
second term is proportional to $g_1 F_1 g_{K} F_{K+1}$. Similar expressions can be derived
for $\Pi_{LL, RR}$. See also \protect\cite{Hirn:2004ze}.}.

Using eqn. (\ref{result}), we can proceed to investigate the asymptotic behavior of $\Pi_{V-A}(Q^2)$ as $Q^2 \to \infty$ in the open moose model by expanding this function in powers of $1/Q^2$
\begin{equation}
\Pi_{V-A}(Q^2) = \sum_{n=1}^{\infty} {c_n\over Q^{2n}}~,
\label{asymptotic}
\end{equation}
using the expansion
\begin{equation}
{1\over {M^2 ( Q^2\cdot {\cal I} +M^2)} } = 
{{\cal I} \over Q^2}\,\left( {{\cal I}\over M^2} - {{\cal I}\over Q^2}
+{M^2 \over Q^4} +\ldots \right)
\end{equation}
The first term yields
\begin{equation}
c_1 = -\,{F^2}+{F^4_1 g^2_1 \over 4}
\left< 1 \left| {1\over {M^2}}\right| K \right>~.
\label{wsrip}
\end{equation}
Computation of the matrix element in eqn. (\ref{wsrip}) from eqn. (\ref{massmatrix})
(by induction on $K$, for example)  yields\footnote{This calculation may be simplified by noting
that we may write $M^2 = G\, V^2\, G$, where $G$ is a diagonal matrix of gauge couplings and
$V^2$ is a matrix of ``vevs.'' In this notation, $|1\rangle$ and $|K\rangle$ are eigenvectors of $G$ with
eigenvalue $g_1$, and the second term in eqn. ({\protect\ref{wsrip}}) reduces to $\langle 1| V^{-2} | K \rangle$.}
\begin{equation}
{F^4_1 g^2_1\over 4}  \left< 1 \left| {1\over {M^2}}\right| K \right> = 
\left( \sum_{k=1}^{K+1} {1\over F^2_k}\right)^{-1}\equiv F^2~,
\label{wsri}
\end{equation}
and therefore $c_1\equiv 0$. Comparing with eqn. (\ref{diffcorrelators}), we see this implies
\begin{equation}
\sum_{\hat{l}}\left({g^2_{\hat{l}V}\over m^2_{\hat{l}}} - {g^2_{\hat{l}A}\over m^2_{\hat{l}}} \right)
= F^2~,
\end{equation}
and therefore the first Weinberg sum rule is satisfied for an arbitrary parity-symmetric open moose. The second term in eqn. (\ref{asymptotic}) yields
\begin{equation}
c_2 = -{\,F^4_1 g^2_1\over 4}  \left< 1 | K \right> \propto \delta_{1K}~,
\end{equation}
and therefore vanishes for all $K>1$. Comparing to eqn. (\ref{diffcorrelators}), we see
that
\begin{equation}
\sum_{\hat{l}} \left( g^2_{\hat{l}V} - g^2_{\hat{l}A} \right) \equiv 0~,
\label{wsrii}
\end{equation}
and the second Weinberg sum rule is automatically satisfied as well.

The fact that the symmetries and spectrum corresponds to what one sees in QCD, and that the
Weinberg sum rules are automatically satisfied for any parity-symmetric
open moose motivates the conjecture that perhaps for some choice of parameters such a model
can be considered a dual to a QCD-like theory \cite{Son:2003et} -- perhaps in the large-$N_C$ limit \cite{tHooft:1974jz}.  That this is unlikely, however, can be seen by examining the behavior of
partial sums of the Weinberg sum rules eqns. (\ref{wsri}) and (\ref{wsrii}). In the case of
a ``flat background'', in which all $F_k$ and $g_k$ are taken to be equal, and for $K$ equal
to 20, the fractional errors of the
partial sums are shown in Figs. \ref{flat1} and \ref{flat2} (the partial sums in the second Weinberg
sum rule are normalized to $g^2_{1V}$ to make them dimensionless). 
The values of the $F$-constants and gauge couplings are set by requiring that $f_\pi$ and
$m_\rho$ be correct. We immediately observe that, unlike QCD in which the Weinberg sum rules are reasonably well satisfied including just the $\rho$ and $A_1$, the sum is off by almost 100\% after inclusion of the first vector and axial-vector mesons. In fact, the fractional error is still almost 50\% after 
including the first {\it ten} vector resonances.

\DOUBLEFIGURE[t]{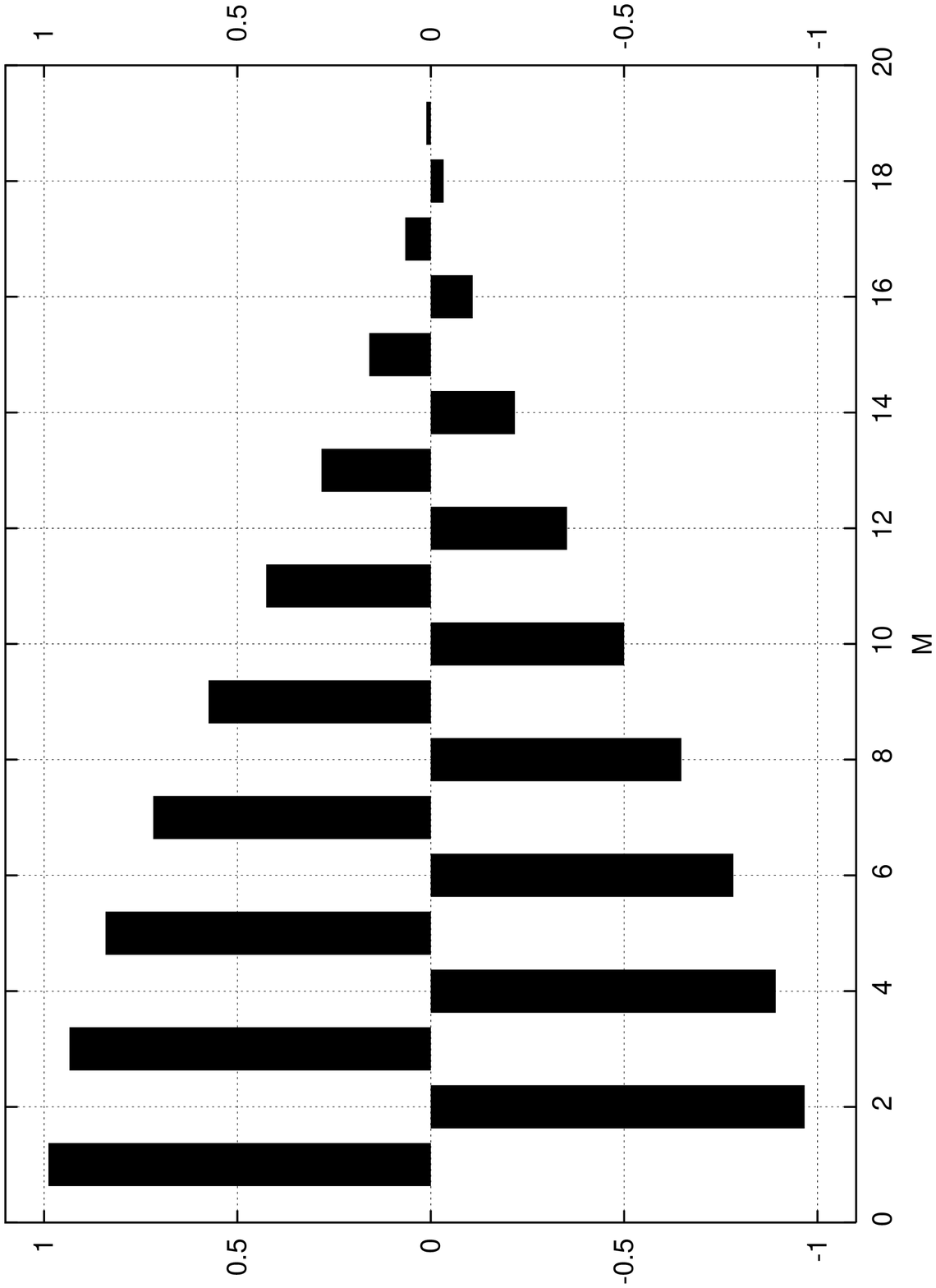, angle=-90,width=0.5\textwidth}{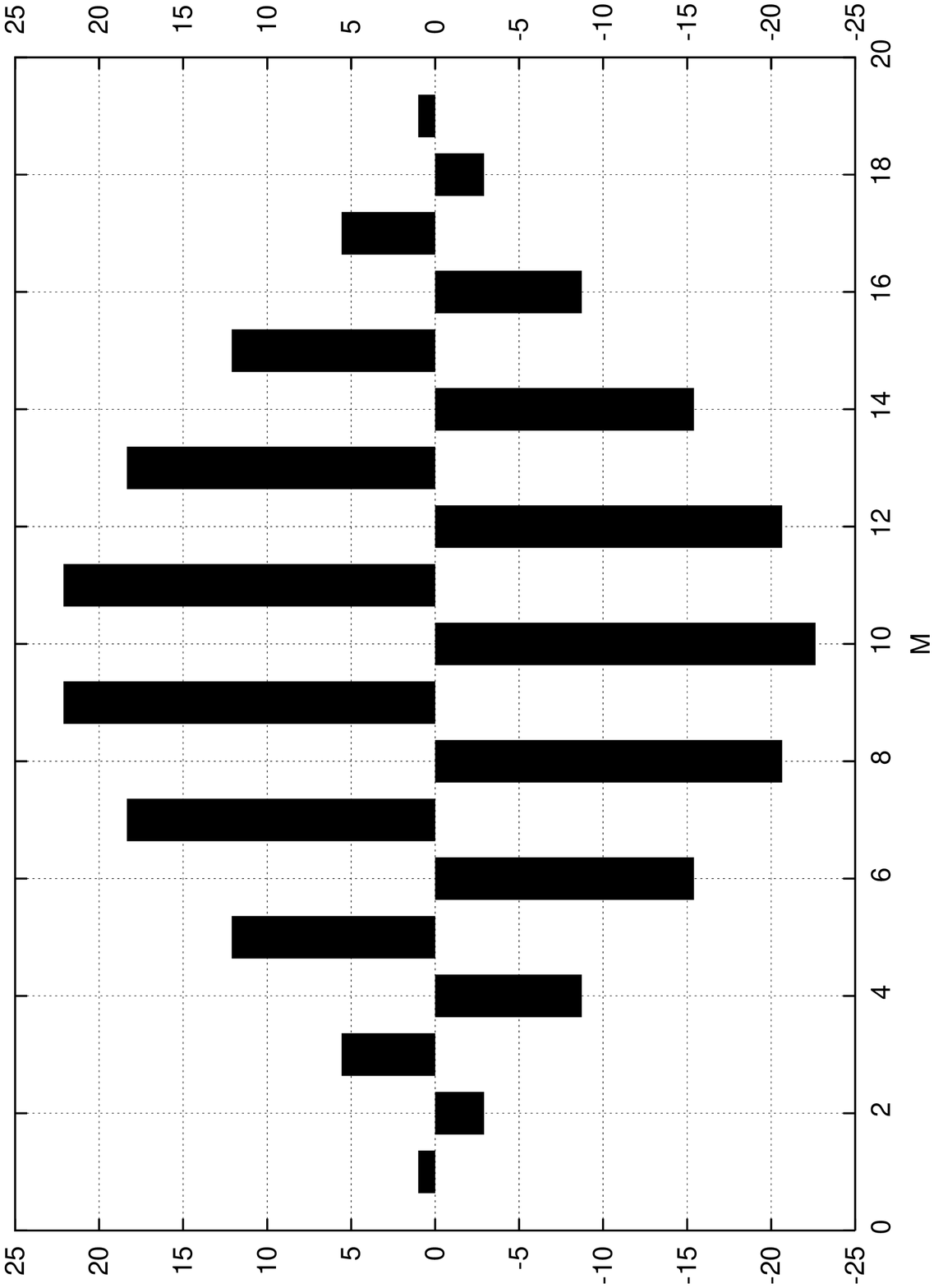, angle=-90,width=0.5\textwidth}
{Fractional error of partial sums of 1st  Weinberg sum rule, $
\left[ {1\over F^{2}_\pi}  \sum_1^M \left( {g^2_{nV} \over m^2_n} - {g^2_{nA}\over m^2_n}\right)-1 \right]$  in the ``flat'' background given twenty lattice sites. Note that more than ten terms are required to reproduce $F^2_\pi$ to 50\%. \label{flat1}}
{Fractional error of partial sums of 2nd  Weinberg sum rule, 
$\left[{1\over g^2_{1V}} \sum_1^M\left( g^2_{nV}-g^2_{nA}\right)\right]$.\label{flat2}}
 
That this behavior is not peculiar to the flat background can be seen in Figs. \ref{cosh1} and
\ref{cosh2}. Here, for illustration, motivated by the ``cosh background'' of \cite{Son:2003et}, 
we take the parameters
\begin{equation}
g_k=g\ \ \ \ \&\ \ \ \ F_k={\Lambda\over g} \cosh\left(-A+(k-1){2A\over K}\right)~,
\end{equation}
with $A=2.0$ and $K=20$. The values of $\Lambda$ and $g$ are, again, set by requiring the
model reproduce $f_\pi$ and $m_\rho$. In this case, the behavior of the partial sums is substantially worse than in the flat case -- in retrospect, this can be understood as arising because the spacing
between the vector and axial-vector mesons grows more quickly in this model \cite{Son:2003et}. Here again we conclude that the model does not behave like QCD.

\DOUBLEFIGURE[t]{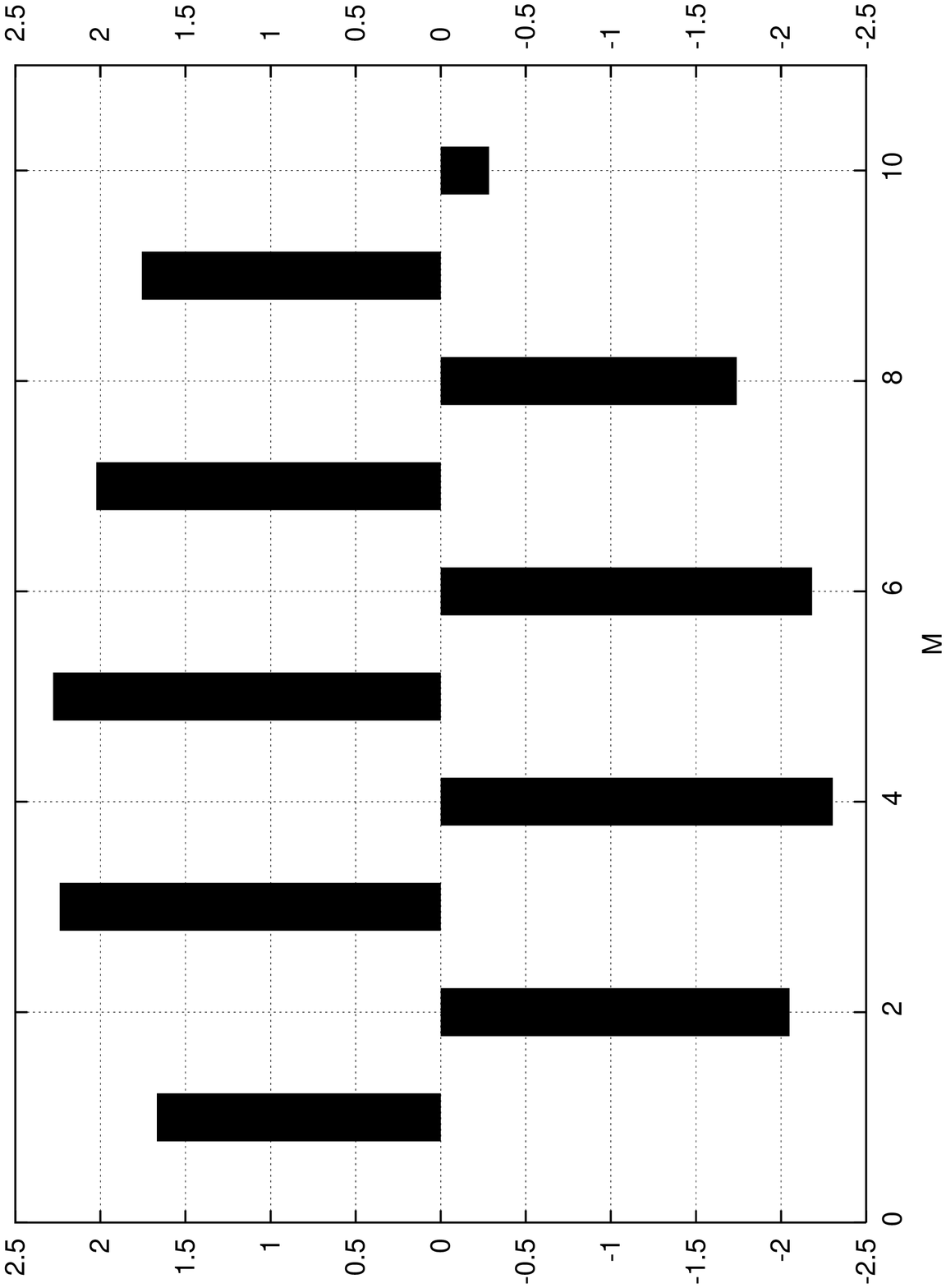, angle=-90,width=0.5\textwidth}{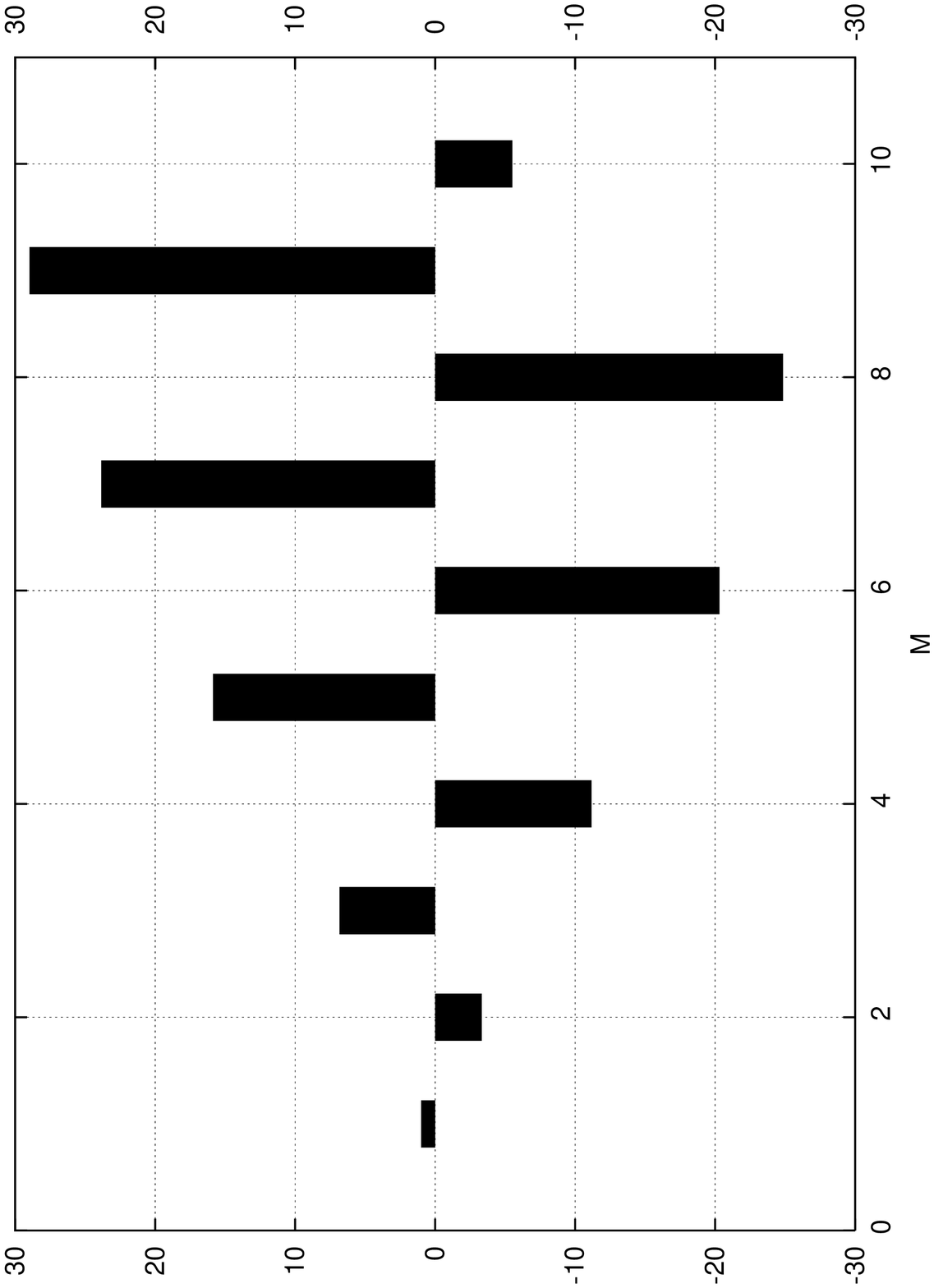, angle=-90,width=0.5\textwidth}
{Same as Fig. \protect\ref{flat1}, for the ``cosh'' background with twenty lattice sites. Only the first ten partial sums are shown; the errors increase for larger values, canceling exactly only when all terms are included. \label{cosh1}}{Same as Fig. \protect\ref{flat2}, for the ``cosh'' background with twenty lattice sites. Only the first ten partial sums are shown; the errors increase for larger values, canceling exactly only when all terms are included. \label{cosh2}}

In fact, we can be substantially more precise in our analysis of the difference between the asymptotic
behavior of the open moose model and that expected in a QCD-like theory. Computing higher order 
terms in the expansion of eqn. (\ref{asymptotic}), we find
\begin{equation}
c_n = (-1)^{n+1}\, {F^4_1 g^2_1\over 4}  \left< 1 \left| (M^2)^{n-2} \right| K \right> ~,
\end{equation}
for $n\ge 2$.
Because of the tridiagonal form of the mass matrix (eqn. (\ref{massmatrix})), we find that
$c_n \equiv 0$ for $n=2, 3, \ldots, K$. Comparing this to eqn. (\ref{diffcorrelators}), we
see that we have a family of generalized sum rules
\begin{equation}
\sum_{\hat{l}} (g^2_{\hat{l}V} - g^2_{\hat{l}A})m^{2j}_{\hat{l}} \equiv 0~,
\label{gsr}
\end{equation}
for $j=0,1,\ldots,K-2$.

The coefficients of the terms in the asymptotic expansion of $\Pi_{V-A}$ can be matched
to the presence or absence of chiral-symmetry breaking condensates 
\cite{Shifman:1979bx, Knecht:1998ts}. In general, 
the constants $c_n$ correspond to the vacuum expectation value of a chiral-symmetry
breaking condensate of dimension $2n$. In QCD-like theories, the asymptotic behavior
\cite{Shifman:1979bx} is expected to be dominated by 
\begin{equation}
c_3=-8\pi^2\left({\alpha_s\over \pi} +{\cal O}(\alpha^2_s)\right)\langle \bar{\psi}\psi\rangle^2~,
\end{equation}
at least in the large-$N_C$ \cite{tHooft:1974jz}  limit.
Therefore, for $K>2$, the open moose does not reproduce the expected behavior. Indeed,
in general the first nonzero expectation value occurs for $c_{K+1}$. It is easy to construct
a perturbative example that realizes this behavior: consider a model in which each of the nonlinear
sigma model fields $\Sigma^k$ are replaced by linear sigma model fields $\Phi^k$
\cite{Bars:1973xk}. In this case,  
$c_{K+1} \propto \langle \Phi^1 \Phi^2 \ldots \Phi^{K+1}\rangle^2$. The absence of low-dimension
chiral symmetry breaking condensates is a manifestation of the ``collective breaking'' incorporated
in these models \cite{Arkani-Hamed:2001nc,Arkani-Hamed:2002pa}.

The generalized sum rules allow for a convenient expression of $\Pi_{V-A}(Q^2)$ in this
model. Consider forming a common denominator for the expression on the right hand side
of eqn. (\ref{diffcorrelators})
\begin{equation}
\Pi_{V-A} (Q^2) = {P(Q^2) \over
Q^2\cdot(Q^2+m^2_{\hat{1}})\ldots (Q^2+m^2_{\hat{K}})}~,
\end{equation}
where, from the form of eqn. (\ref{diffcorrelators}), we see that $P(Q^2)$ is a polynomial
in $Q^2$ at most of order $K$. However, as discussed above, the generalized sum-rules of eqn. (\ref{gsr}) imply that $\Pi_{V-A}(Q^2)$ must fall like $1/(Q^2)^{K+1}$ for large $Q^2$. {\it Therefore}
$P(Q^2)$ {\it must be a constant.} We may evaluate this constant by requiring that the pion pole
have the correct residue, and hence we derive
\begin{equation}
\Pi_{V-A}(Q^2) = -\, {F^2 \over Q^2}\, \prod_{\hat{l}} {m^2_{\hat{l}} \over (Q^2+m^2_{\hat{l}})}~.
\label{greatresult}
\end{equation}
The residues of the poles at the masses $m^2_{\hat{l}}$ provide an expression for the vector
meson decay constants in terms of the masses and $F$
\begin{equation}
g^2_{\hat{l}V}-g^2_{\hat{l}A} =  m^2_{\hat{l}} \, F^2 \prod_{\hat{k}\neq \hat{l}}
{m^2_{\hat{k}} \over m^2_{\hat{k}}-m^2_{\hat{l}}}~,
\label{vectorconstants}
\end{equation}
``solving'' the generalized sum rules (eqn. (\ref{gsr})) and completing
the program of  \cite{Weinberg:1967kj,Knecht:1998ts} in this model.

We may now immediately compute the analogs of the $S$ ($L_{10}$) parameter and the
$\pi^+$ -- $\pi^0$ mass difference in QCD. $S$  \cite{Gasser:1984yg,Gasser:1985gg,Peskin:1990zt,Holdom:1990tc,Golden:1991ig,Peskin:1992sw}
is calculated to be
\begin{equation}
S = 4\pi\, {d \over dQ^2} \left[Q^2\, \Pi_{V-A}(Q^2)\right]_{Q^2=0}
= 4\pi F^2\, \left(\sum_{\hat{l}} {1\over m^2_{\hat{l}}}\right)~,
\label{sparameter}
\end{equation}
and is manifestly positive in any open moose model.\footnote{An alternative, manifestly positive,
expression for $S$ is given in \protect\cite{Hirn:2004ze}.}  We note that this expression
implies that any ``Higgsless model''  \cite{Csaki:2003dt,Csaki:2003zu,Barbieri:2003pr,Foadi:2003xa}
based on the continuum limit of a linear moose will have a positive $S$ parameter, 
regardless\footnote{In this paper, we have calculated $L_{10}$ in the effective chiral $SU(2)_L \times
SU(2)_R$ theory. As such, our results are only accurate to lowest nontrivial order in the
electroweak couplings $g$ or $g'$. It has been shown 
\protect\cite{Cacciapaglia:2004jz} that, beyond this approximation, a brane kinetic energy term of the correct type can give rise to a negative contribution
to $S$.}  of the warping or position-dependent gauge-coupling chosen. In any unitary theory \cite{SekharChivukula:2001hz, Chivukula:2002ej}, we expect the mass of the lightest
vector $m_{\hat{1}}$ to be less than $\sqrt{8 \pi} v$ ($v\approx 246$ GeV) -- the scale at
which $WW$ spin-0 isospin-0 elastic scattering would violate unitarity in the 
standard model in the absence of a higgs
boson \cite{Dicus:1973vj,Cornwall:1973tb, Cornwall:1974km,Lee:1977yc, Lee:1977eg,Veltman:1977rt}.
Evaluating eqn. (\ref{sparameter}), with $F=v$, we see that we expect $S$ to be of order one-half  or larger.

Similarly, the pion mass-squared
difference \cite{Das:1967it,Harada:2003xa} may be written
\begin{equation}
\Delta m^2_\pi = {3 \alpha_{em} \over 4\pi F^2}\,
\int^\infty_0 dQ^2\,(-Q^2)\,\Pi_{V-A}(Q^2)~,
\end{equation}
and, from eqn. (\ref{greatresult}), is again manifestly positive as expected for a vector-like
theory \cite{Vafa:1984tf}.

 \acknowledgments

We thank Nick Evans, Hong-Jian He, Elizabeth H. Simmons, John Terning, and Koichi Yamawaki for
discussions, and Johannes Hirn and Jan Stern of informing us of their overlapping results.
M.K. acknowledges support by the 21st Century COE Program of Nagoya University 
provided by JSPS (15COEG01), and thanks the Michigan State University high-energy theory group for their hospitality during the completion of this work. M.T.'s work is supported in part by the JSPS Grant-in-Aid for the Scientific Research No.14340072.

\appendix

\section{Appendix}

The analogs of the $S$ ($L_{10}$) parameter and the $\pi^+$ -- $\pi^0$ mass difference in QCD  may also be expressed as matrix elements in this model.
In the pole approximation \cite{Gasser:1984yg,Peskin:1992sw},
\begin{equation}
S=4\pi \sum_{\hat{l}} {{g^2_{\hat{l}V} - g^2_{\hat{l}A}}\over m^4_{\hat{l}}} = \pi\, F^4_1 g^2_1
  \left< 1 \left| {1\over M^4}  \right| K \right> ~,
\label{ssimple}
\end{equation}
where the last equality follows from the derivation given above. Evaluating this expression for
the ``flat background'' mass matrix, we find
\begin{equation}
S= {8\pi \over 3}\, {1\over g^2}\, {K(K+2)\over K+1}~,
\end{equation}
in agreement with \cite{Foadi:2003xa}. The derivation of eqn. (\ref{ssimple}) relies only on the two global
symmetry groups coupling to only one gauge group each -- not on the linearity of the moose.
It is interesting to speculate that a $M^2$ matrix can be found for which $S\equiv 0$, but
for which chiral symmetry breaking still takes place\footnote{We thank E. H. Simmons for discussions
of this point.} -- such a theory may provide the basis for a phenomenologically interesting theory of dynamical electroweak symmetry breaking.

Similarly, the pion mass-squared difference can be
written \cite{Das:1967it}
\begin{equation}
\Delta m^2_\pi = -\, {3\alpha_{em}\over 4\pi F^2}\, \sum_{\hat{l}}(g^2_{\hat{l}V}-g^2_{\hat{l}A})
\log m^2_{\hat{l}}~,
\end{equation}
where, because of the 2nd Weinberg sum rule, the expression is finite. The methods above
imply that in the open moose model, this may be formally written as
\begin{equation}
\Delta m^2_\pi = -\, {3\alpha_{em}F^4_1 g^2_1\over 16 \pi F^2}
\left< 1 \left| \log M^2 \right| K \right> ~.
\end{equation}
%



\providecommand{\href}[2]{#2}\begingroup\raggedright\endgroup

\end{document}